\documentclass[aps,preprint,prd,showpacs,nofootinbib]{revtex4}
\usepackage{amsmath}
\usepackage{graphicx}
\usepackage{subfigure}
\usepackage{multirow}
\usepackage{bm}
\usepackage{amssymb}
\usepackage{mathtools}
\usepackage{enumerate}
\usepackage{color}
\usepackage[colorlinks,linkcolor=magenta,anchorcolor=cyan,citecolor=blue]{hyperref}
\usepackage{soul}

\def\be{\begin{equation}}
\def\ee{\end{equation}}

\begin{document}
\title{Improved constraint on primordial gravitational waves in light of
the Hubble tension and BICEP/Keck}

\author{Gen Ye$^{1}$\footnote{yegen14@mails.ucas.ac.cn}}
\author{Yun-Song Piao$^{1,2,3,4}$\footnote{yspiao@ucas.ac.cn}}

\affiliation{$^1$ School of Physics, University of Chinese Academy of
    Sciences, Beijing 100049, China}

\affiliation{$^2$ School of Fundamental Physics and Mathematical
    Sciences, Hangzhou Institute for Advanced Study, UCAS, Hangzhou
    310024, China}

\affiliation{$^3$ International Center for Theoretical Physics
    Asia-Pacific, Beijing/Hangzhou, China}

\affiliation{$^4$ Institute of Theoretical Physics, Chinese
    Academy of Sciences, P.O. Box 2735, Beijing 100190, China}

\begin{abstract}

The Hubble tension that the standard $\Lambda$CDM model is
suffering from can be resolved with pre-recombination early dark
energy. We present the first constraint on the tensor-to-scalar
ratio $r$ in corresponding Hubble-tension-free cosmologies using
the most recent BICEP/Keck cosmic microwave background (CMB)
B-mode polarization data. We find, combining BICEP/Keck with
Planck18 CMB and baryon acoustic oscillation data, that the models
with larger Hubble constant $H_0$ will have tighter upper bound on $r$,
and resolution $H_0\sim73$ km/s/Mpc of the Hubble tension tightens
the upper bound to $r<0.028\ (95\%\text{C.L.})$, 25\% tighter than
the $\Lambda$CDM constraint $r<0.036$. We clarify the origin of
this tightening bound.

\end{abstract}
\maketitle
\section{Introduction}\label{sec:intro}

Inflation is the current paradigm of early universe
\cite{Guth:1980zm,Linde:1981mu,Albrecht:1982wi,Starobinsky:1980te}.
It predicts nearly scale-invariant scalar perturbation, which is
consistent with the cosmic microwave background (CMB)
observations, as well as the constraint on primordial gravitational waves (GWs).
The discovery of primordial GWs will solidify our confidence that
inflation has ever happened. The primordial GWs will source B-mode polarization in the CMB
\cite{Seljak:1996ti,Kamionkowski:1996zd,Seljak:1996gy}, which is currently the most promising way to search for the primordial
GWs.

Based on the standard $\Lambda$CDM model, combining Planck18 and
BICEP/Keck15 data the Planck collaboration has put the constraint
on the tensor-to-scalar ratio, $r<0.066$ (95\% C.L.)
\cite{Planck:2018vyg}. Recently, combining Planck18, baryon
acoustic oscillations (BAO) and latest BICEP/Keck18 data the
BICEP/Keck collaboration has lowered the upper bound to $r<0.036$
(95\% C.L.) \cite{BICEP:2021xfz}\footnote{A slightly tighter bound $r<0.032$ is
obtained using the new Planck PR4 data \cite{Tristram:2021tvh}.}.
However, it is well-known that the $\Lambda$CDM model suffers the Hubble tension, i.e. the locally measured value of
current expansion rate reported by the SH0ES collaboration is
$H_0\sim 73.04\pm1.04$km/s/Mpc \cite{Riess:2021jrx}, in stark
($\sim5\sigma$) tension with $H_0=67.37\pm0.54$km/s/Mpc
\cite{Planck:2018vyg} inferred by the Planck collaboration
assuming $\Lambda$CDM.

Though the possibility of some unknown systematics in data causing
this tension \cite{Freedman:2021ahq} cannot be ruled out, the
Hubble tension is actually becoming a pointer to new physics
beyond $\Lambda$CDM, see
e.g.\cite{DiValentino:2021izs,Perivolaropoulos:2021jda} for
reviews. The inclusion of early dark energy (EDE)
\cite{Poulin:2018cxd}, see also \cite{Agrawal:2019lmo,
Alexander:2019rsc,Lin:2019qug,Niedermann:2019olb,Sakstein:2019fmf,Ye:2020btb,Braglia:2020bym,Karwal:2021vpk,McDonough:2021pdg},
has proved to be a promising route of resolving the Hubble
tension. In Hubble-tension-free EDE cosmologies, the bestfit
values of cosmological parameters have shifted notably in
correlation with the increment in $H_0$ \cite{Ye:2021nej}, see
also
\cite{Hill:2020osr,Ye:2020oix,Pogosian:2020ded,Vagnozzi:2021gjh}.
The parameter shifts can serve as tests of corresponding EDE
models, which have been confronted with large scale structure data
\cite{Hill:2020osr,Ivanov:2020ril,DAmico:2020ods} and high-$l$ CMB
data from ground based experiments
\cite{Chudaykin:2020acu,Chudaykin:2020igl,Jiang:2021bab,Poulin:2021bjr,Hill:2021yec,
LaPosta:2021pgm}, respectively.

\begin{figure}
\includegraphics[width=0.8\linewidth]{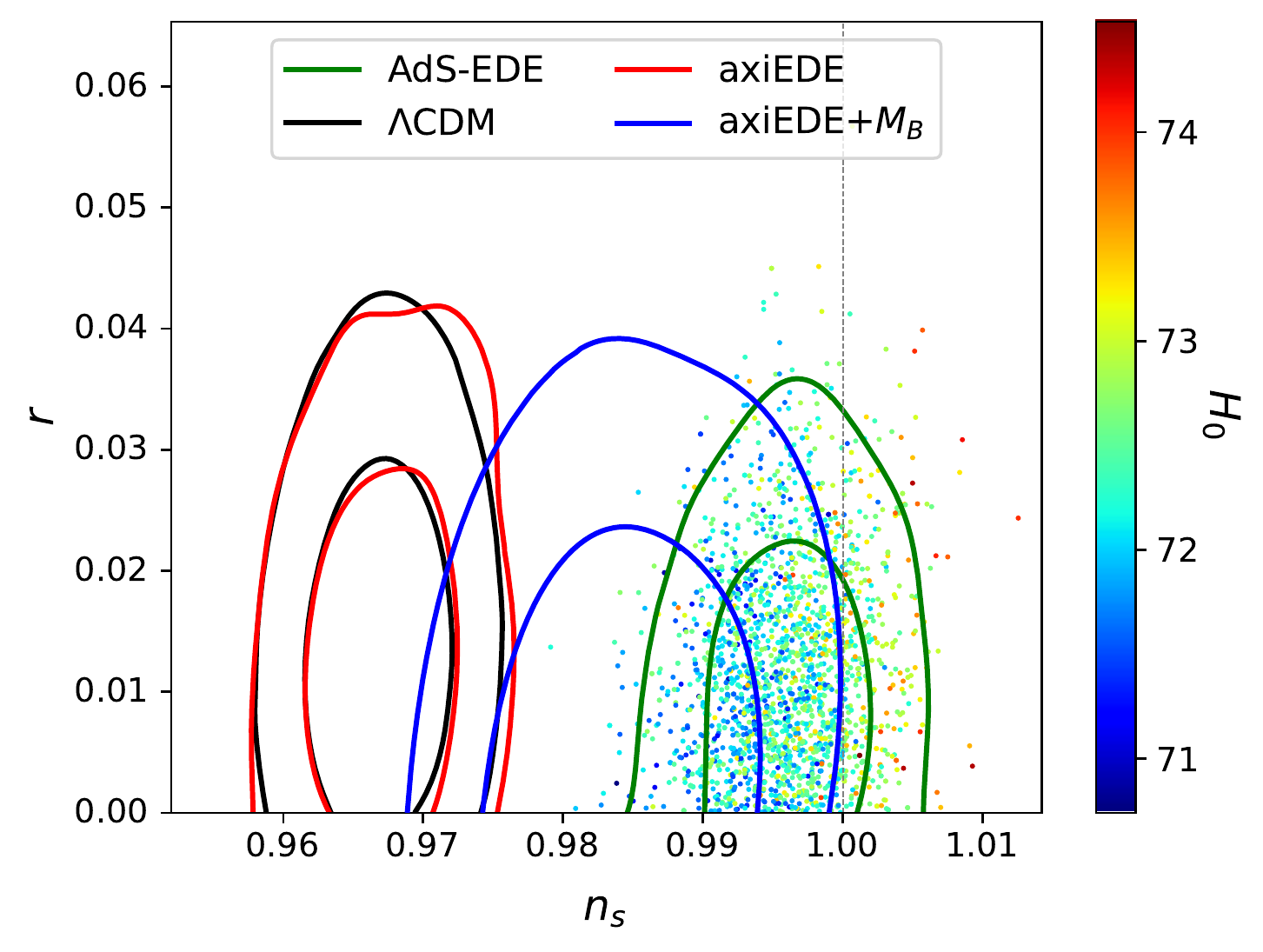}
\caption{68\% and 95\% C.L. contour plot of the tensor-to-scalar
ratio $r$ versus the primordial scalar spectrum tilt $n_s$, with a
color coding for $H_0$. All EDE results include the dataset
P18+BK18+BAO+SN, while axiEDE+$M_B$ additionally includes the
SH0ES result as a Gaussian prior on the absolute magnitude
calibration $M_B$ \cite{Camarena:2021jlr,Efstathiou:2021ocp}. The
$\Lambda$CDM result is from Ref.\cite{BICEP:2021xfz}
with the dataset P18+BK18+BAO. Generally, the EDE models with
larger $H_0$ have larger $n_s$ and tighter upper bound on
$r$. Resolution of the Hubble tension ($H_0\sim73$km/s/Mpc)
roughly corresponds to $n_s\approx1$ Ref.\cite{Ye:2021nej} and a
25\% tighter $r$ upper bound, see (\ref{r}).}
    \label{r_h_ns}
\end{figure}

The amplitude $A_s$, tilt $n_s$ of primordial scalar perturbations
and the tensor-to-scalar ratio $r$ set the initial condition of
CMB. They are converted into the observed CMB anisotropies
through perturbation evolution described by a certain cosmological
model. Thus it is to be expected that any constraints on the
relevant parameters acquired assuming $\Lambda$CDM will get
modified in the new cosmologies. It has been found in
\cite{Ye:2021nej} that the shift of primordial scalar spectral
index scales as 
\begin{equation}
{\delta n_s}\simeq 0.4{\delta H_0\over
H_0},\label{deltans}
\end{equation} and so the Hubble-tension-free cosmologies
actually suggests a scale-invariant Harrison-Zeldovich primordial
scalar spectrum, i.e. $n_s=1$ for $H_0\sim73$km/s/Mpc. $n_s=1$
have profound implication on inflation and primordial Universe,
see e.g. recent Refs.\cite{Takahashi:2021bti,DAmico:2021zdd}.

In view of the inevitability of Hubble tension in $\Lambda$CDM \cite{Verde:2019ivm}, it is significant and also imperative
to constrain $r$ in Hubble-tension-free cosmologies,
e.g.\cite{Poulin:2018cxd,Ye:2020btb}, using the most recent
BICEP/Keck B-mode polarization data. In this Letter, we will
present the first constraint on $r$ in corresponding cosmologies.
We find that the models predicting larger $H_0$ will have larger
$n_s$ and tighter upper bound on $r$, and resolution
$H_0\sim73$km/s/Mpc of the Hubble tension tightens the upper bound
on $r$ to 
\begin{equation}\label{r}
	r<0.028 \ (95\%\text{C.L.}),
\end{equation}
25\% tighter than the $\Lambda$CDM constraint
$r<0.036$ reported by the BICEP/Keck collaboration
\cite{BICEP:2021xfz}, see Fig.\ref{r_h_ns}. We clarify the origin of this tightening
bound\footnote{Using the BICEP/Keck15 data \cite{BICEP2:2018kqh},
Ref.\cite{Ye:2021nej} found the upper bounds on $r$ in EDE to be
similar to that in $\Lambda$CDM. However, the situation is different with sufficiently
precise B-mode data.}, and comment on the
possibility to relax it.

%The paper is outlined as follows. In section-\ref{sec:model} we
%review the cosmological models ,datasets and analysis methods
%used. Our results are presented in section-\ref{sec:results} as
%well as their corresponding physical discussion and implication.
%We conclude the paper in section-\ref{sec:conclusion}. Technical
%details and full numerical results are gathered in the Appendix.

\section{Model and data}\label{sec:model}

As concrete examples of Hubble-tension-free cosmologies, we limit
ourself to the EDE, which must be non-negligible only for a short
epoch decades before recombination and can be implemented as a
canonical scalar field $\phi$ with a potential $V(\phi)$. The EDE
models we consider will be: axion-like EDE \cite{Poulin:2018cxd}
with an oscillating potential
$V(\phi)=m^2f_a^2(1-\cos(\phi/f_a))^n$ and $n=3$ (denoted as
axiEDE for simplicity), and AdS-EDE \cite{Ye:2020btb} with a
rolling potential $V(\phi)=V_0({\phi\over M_p})^{4}-V_{ads}$ glued
to a cosmological constant $V(\phi)=const.>0$ at
${\phi}=(\frac{V_{ads}}{V_0})^{1/4}M_p$, where $V_{ads}$ is the
depth of anti-de Sitter (AdS) well. The evolution of Universe
after recombination is $\Lambda$CDM-like, see also
e.g.\cite{Wang:2022jpo}.

We use modified versions\footnote{The corresponding
cosmological codes are available at: axiEDE
(\url{https://github.com/PoulinV/AxiCLASS}) and AdS-EDE
(\url{https://github.com/genye00/class_multiscf}).} of CLASS
\cite{Lesgourgues:2011re,Blas:2011rf} to compute cosmology
and the MontePython-3.4 sampler
\cite{Audren:2012wb,Brinckmann:2018cvx} to perform Monte Carlo
Markov Chain (MCMC) analysis. In addition to the six $\Lambda$CDM
parameters $\{\omega_{b},\omega_{cdm},H_0,
\ln10^{10}A_s,n_s,\tau_{reion}\}$, we vary two additional MCMC
parameters $\{\ln(1+z_c),f_{ede}\}$, with $z_c$ being the redshift
at which the field $\phi$ starts rolling and $f_{ede}$ the energy
fraction of EDE at $z_c$, for both EDE models. The axiEDE model
varies yet one more MCMC parameter $\Theta_i\equiv\phi_i/f_a$, the
initial position of the field. We set $n_T=0$ following BK18
\cite{BICEP:2021xfz}.

Our datasets include:
\begin{itemize}
\item \textbf{P18:} Planck 2018 high-$l$ TTTEEE, low-$l$ TT and
EEBB likelihoods as well as Planck lensing \cite{Planck:2019nip}.
\item \textbf{BK18:} The most recent CMB B-mode
polarization data from BICEP/Keck 2018 \cite{BICEP:2021xfz}.
    \item \textbf{BAO:} Post-reconstructed BAO measurements from 6dF \cite{Beutler:2011hx}, MGS \cite{Ross:2014qpa} and BOSS DR12 \cite{BOSS:2016wmc}.
\item \textbf{SN:} the Pantheon dataset, with a single nuisance
parameter $M_B$, calibrating the absolute magnitude of the
supernovas \cite{Scolnic:2017caz}.
%\item \textbf{$\mathbf{M_B}$:}
%the SH0ES measurement of $H_0$ as a Gaussian prior
%$M_B=-19.214\pm0.037$ mag\footnote{The most recent SH0ES result
%$H_0=73.04\pm1.04$ km/s/Mpc corresponds to $M_B=-19.253\pm0.027$
%mag \cite{Riess:2021jrx} in the new Pantheon+ sample
%\cite{Scolnic:2021amr,Brout:2021mpj}, which was released during
%the preparation of this paper. Since the new Pantheon+ sample is
%compatible with the Pantheon dataset used in this paper, inclusion
%of new data should not change our qualitative results.}.
\end{itemize}
In the following we also include the standard $\Lambda$CDM model for reference. However, we do not redo the MCMC analysis for the
$\Lambda$CDM model but directly use the $\Lambda$CDM chains from
BK18 \footnote{Available at
\url{http://bicepkeck.org/bk18_2021_release.html}}
\cite{BICEP:2021xfz} (P18+BK18+BAO).

\section{Results and discussion}\label{sec:results}
\begin{figure}
\includegraphics[width=\linewidth]{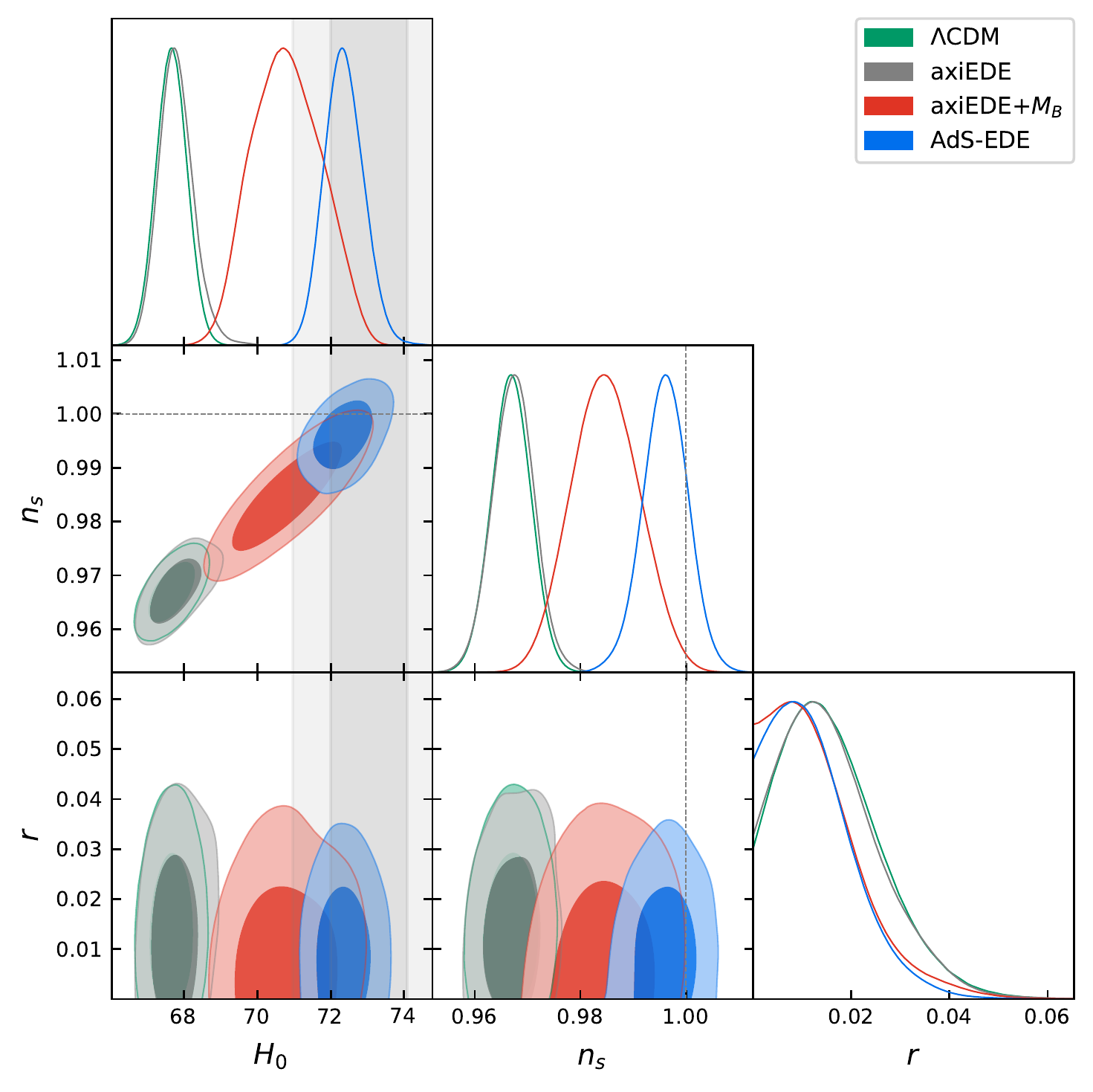}
\caption{68\% and 95\% posterior distribution of the $\Lambda$CDM,
axiEDE(w/ and w/o $M_B$ prior) and AdS-EDE models. The
$\Lambda$CDM contours are produced using the publicly available
BK18 chains. Gray bands represent the 1$\sigma$ and 2$\sigma$
regions of the SH0ES measurement $H_0=73.04\pm1.04$ km/s/Mpc.
$n_s=1$ is marked by dotted lines.}
    \label{triangle-lite}
\end{figure}

\begin{table}
    \begin{tabular}{|c|c||c|c|c|}
        \hline
        Parameter&$\Lambda$CDM&axiEDE (w/o $M_B$ prior)&axiEDE (w/ $M_B$ prior)&AdS-EDE\\
        \hline
        $100\omega_b$&$2.242(2.231)\pm0.013$ &$2.239(2.276)^{+0.013}_{-0.016}$ &$2.291(2.281)\pm 0.023$ &$2.334(2.343)\pm 0.017$ \\
        $\omega_{cdm}$&$0.1193(0.1195)\pm0.0009$ &$0.11983(0.1248)^{+0.0009}_{-0.0013}$ &$0.1270(0.1302)\pm 0.0035$ &$0.1342(0.1327)\pm 0.0018$ \\
        $\tau_{reio }$&$0.0565(0.0526)\pm0.0073$&$0.0562(0.0642)^{+0.0067}_{-0.0075}$ &$0.0542(0.0551)\pm 0.0074$ &$0.0535(0.0552)\pm 0.0077$ \\
        $H_0$&$67.66(67.57)\pm0.42$& $67.78(70.02)^{+0.41}_{-0.50}$&$70.82(71.83)^{+0.90}_{-1.0}$ &$72.36(72.44)^{+0.49}_{-0.56}$ \\
        $\ln10^{10}A_{s}$&$3.048(3.036)\pm0.014$&$3.045(3.077)\pm 0.015$ &$3.091(3.073)^{+0.019}_{-0.015}$ &$3.070(3.071)\pm 0.015$ \\
        $n_s$&$0.9669(0.9658)\pm0.0037$ &$0.9673(0.9868)\pm 0.0039$ &$0.9848(0.9907)\pm 0.0063$ &$0.9961(0.9978)\pm 0.0043$\\
        $r$&$<0.035(0.015)$&$<0.035(0.011)$ &$< 0.031(0.011)$\ &$<0.028(0.008)$ \\
        $f_{ede}$&--&$<0.024(0.070)$ &$0.075(0.116)^{+0.019}_{-0.035}$ &$0.1120(0.1062)^{+0.0040}_{-0.0076}$ \\
        $\ln(1+z_c)$&--&$8.50(8.81)\pm 0.63$ &$8.090(8.258)^{+0.083}_{-0.18}$ &$8.189(8.178)\pm 0.079$ \\
        $\Theta_i$&--&$1.68(2.97)^{+1.3}_{-0.73}$ &$1.65(2.76)^{+0.30}_{-0.57}$ &-- \\
        \hline
        $\Omega_m$&$0.3111(0.3119)\pm0.0057$&$0.3110(0.3022)\pm 0.0054$ &$0.3002(0.2978)^{+0.0048}_{-0.0056}$ &$0.3021(0.2976)\pm 0.0054$ \\
        $S_8$&$0.826(0.823)\pm 0.010$&$0.825(0.832)\pm 0.010$ &$0.843(0.836)^{+0.013}_{-0.010}$ &$0.858(0.849)\pm 0.011$ \\
        \hline
    \end{tabular}
\caption{Bestfit (in parenthesis) and 68\% C.L. marginalized
constraints (for one-sided bounds the 95\% result is given) on the
cosmological parameters of the $\Lambda$CDM and EDE models. The
$\Lambda$CDM constraints are calculated from the publicly
available BK18 chains with the dataset P18+BK18+BAO. The axiEDE
and AdS-EDE results are obtained using the dataset
P18+BK18+BAO+SN, and axiEDE with $M_B$ prior also include the
SH0ES result as a Gaussian prior on the absolute magnitude
calibration $M_B$.}
    \label{partab}
\end{table}

\begin{figure}
\includegraphics[width=\linewidth]{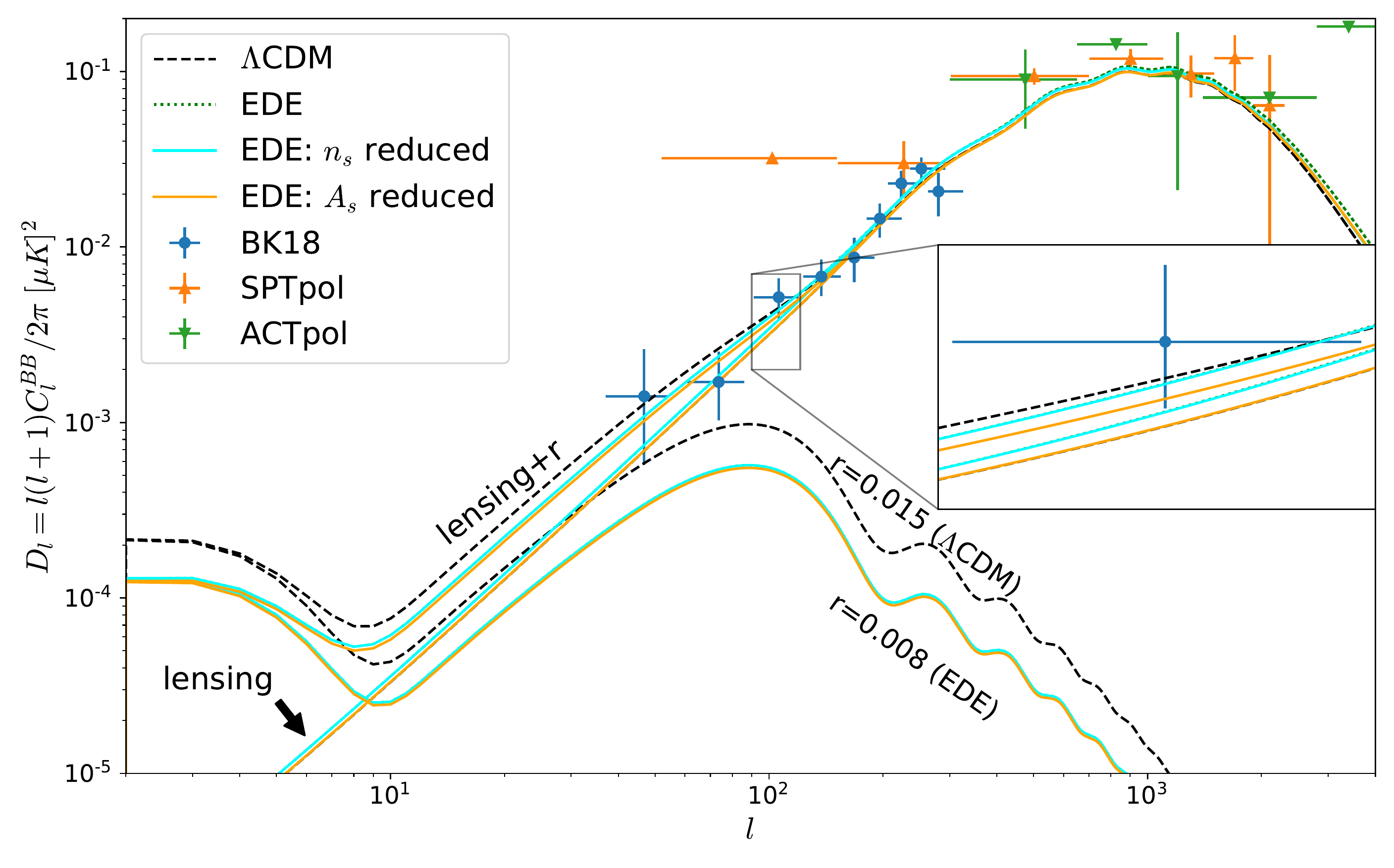}
\caption{$C_{l,tot}^{BB}$, $C_{l,lensing}^{BB}$ and
$C_{l,tensor}^{BB}$ with different color and line style for the
bestfit $\Lambda$CDM (dashed black), bestfit AdS-EDE (dotted
green) and bestfit AdS-EDE but with $A_s$ (solid orange) or $n_s$
(solid cyan) set to their $\Lambda$CDM bestfit values. Points with
error bars are binned BK18 \cite{BICEP:2021xfz}, SPT \cite{SPT:2019nip} and ACT \cite{ACT:2020frw} data points. The $n_s$-reduced AdS-EDE
lines are nearly identical to those of AdS-EDE bestfit thus the
dotted green lines overlap with the cyan ones and are barely
visible.}
    \label{clbb}
\end{figure}

The MCMC posterior results of $\{H_0, n_s, r\}$ for the
$\Lambda$CDM and EDE models are presented in Table.\ref{partab}
and Fig.\ref{triangle-lite}, see appendix-\ref{apdx:mcmc} for the
results of all relevant parameters. Here, what we intend to
discuss is the tighter upper bound on the tensor-to-scalar ratio
$r$. In Fig.\ref{triangle-lite}, we see that in EDE models the
upper bound on $r$ becomes tighter as $H_0$ increases, and with AdS-EDE
($H_0=72.36^{+0.49}_{-0 56}$km/s/Mpc) we have the tightest bound
$r<0.028$, roughly 25\% lower than the $\Lambda$CDM bound. The
origin of this result is quite complicated, which we will clarify
step by step.

We plot the bestfit total B-mode power spectra $C_{l,tot}^{BB}$, the separate contribution from lensing B-mode $C^{BB}_{l,lensing}$ and tensor $C^{BB}_{l,tensor}$, as well as
binned data points in Fig.\ref{clbb}. The difference in $r$ is obvious and the $\Lambda$CDM
(dashed black) and AdS-EDE lines for $C_{l,tot}^{BB}$ are easily distinguishable in Fig.\ref{clbb}, despite both
fit the BK18 data equally well according to Table.\ref{chi2}.
Reducing $n_s$ in AdS-EDE to its $\Lambda$CDM bestfit value (solid
cyan) yields nearly identical results to the AdS-EDE bestfit
(dotted green) in Fig.\ref{clbb}, \textbf{thus $n_s$ is not
directly relevant to the change in the upper bound of $r$}.

%Reducing $A_s$ to its $\Lambda$CDM value pushes the lensing
%$C_{l,lensing}^{BB}$ to be closer to the $\Lambda$CDM line but
%further increases the discrepancy between the tensor B-modes since
%$A_T=rA_s$ is further decreased.

\begin{figure}
\includegraphics[width=0.8\linewidth]{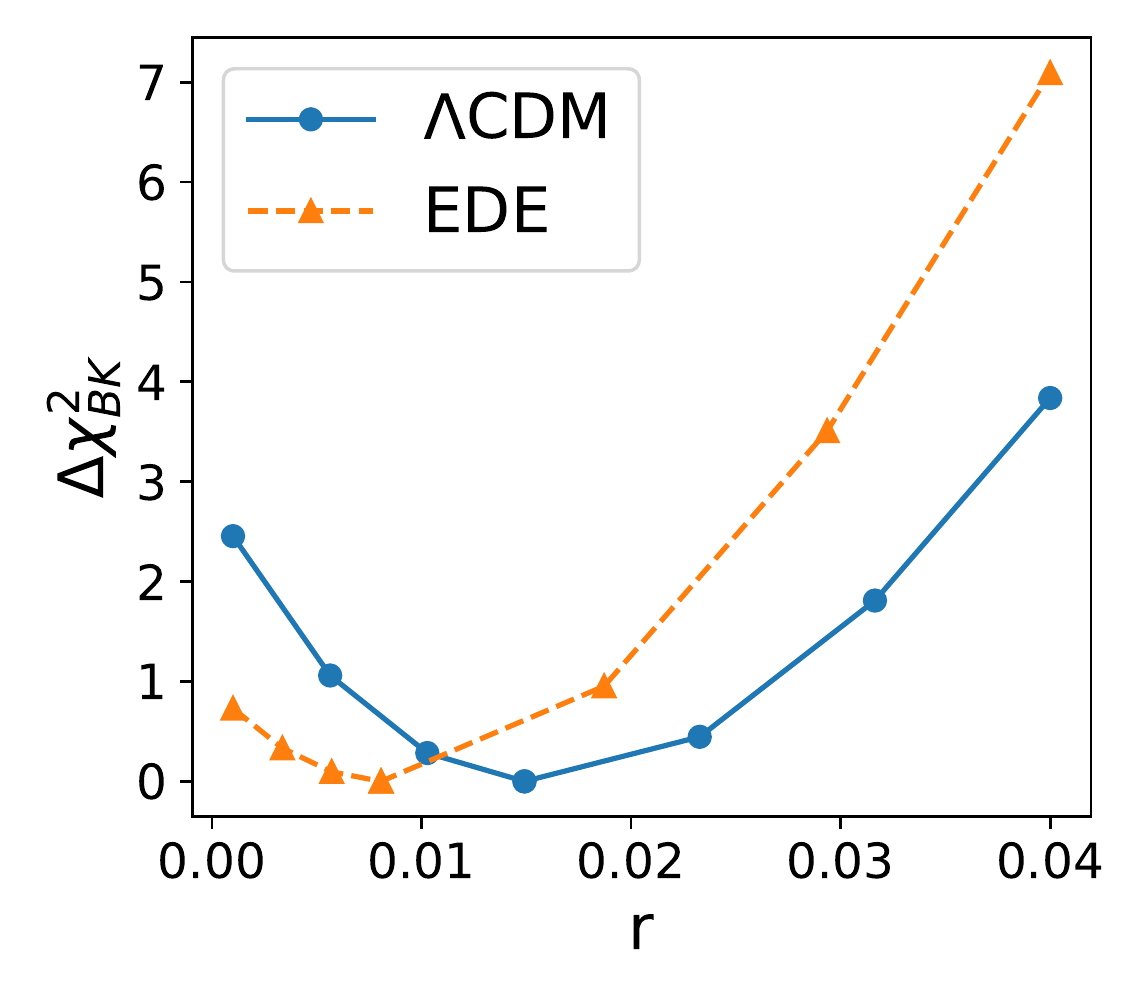}
\caption{Response of total BK18 $\chi^2$ to the variation in $r$ in
$\Lambda$CDM and AdS-EDE. $\chi^2$ at each point is calculated by
varying $r$ with all other parameters, including nuisance, fixed
to their bestfit values. The $y$-axis plots
$\Delta\chi^2=\chi^2-\chi^2_{bestfit}$ for the BK18 dataset.}
    \label{r-chi2}
\end{figure}

\begin{figure}
\subfigure[$\Lambda$CDM]{\includegraphics[width=0.48\linewidth]{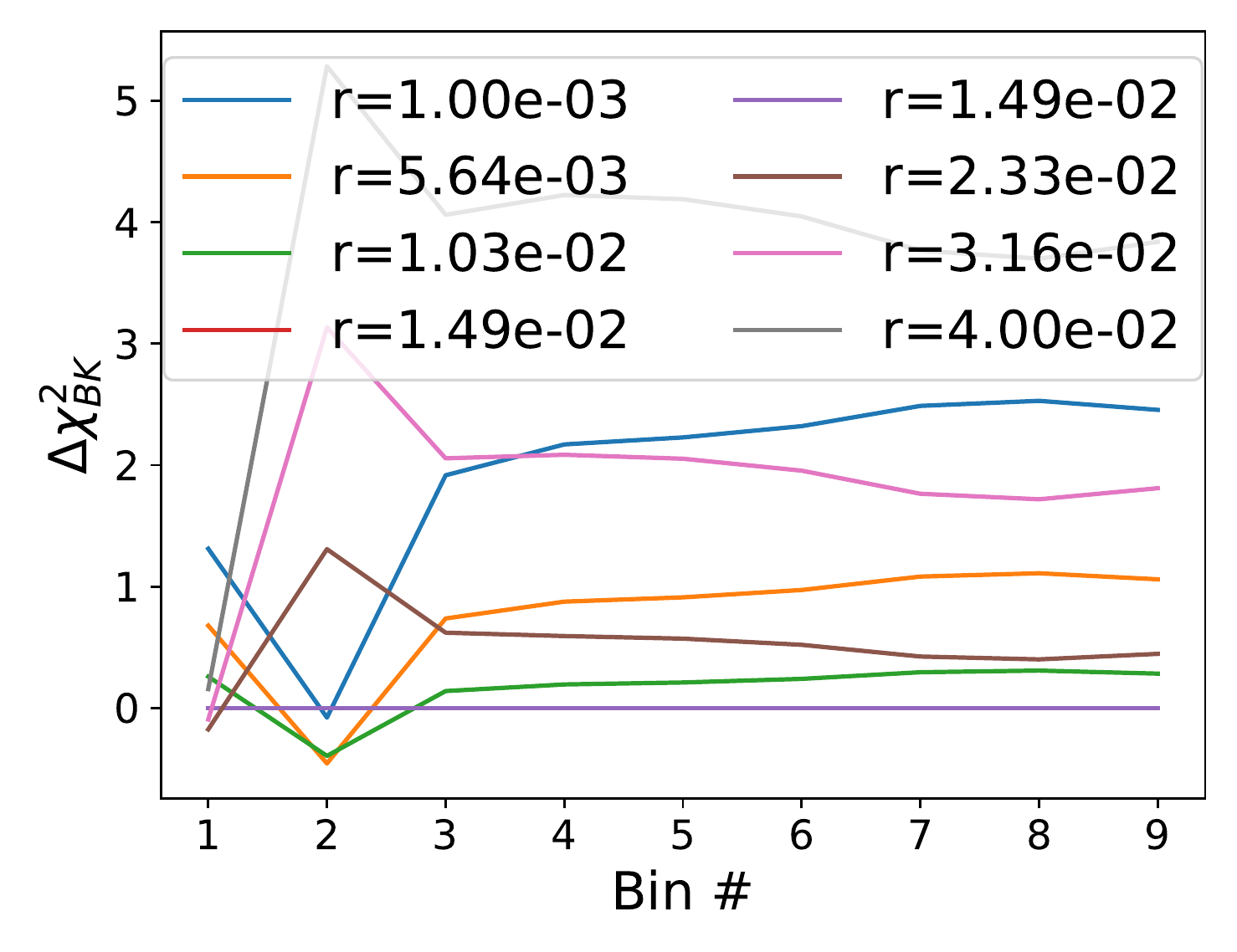}}
\subfigure[AdS-EDE]{\includegraphics[width=0.48\linewidth]{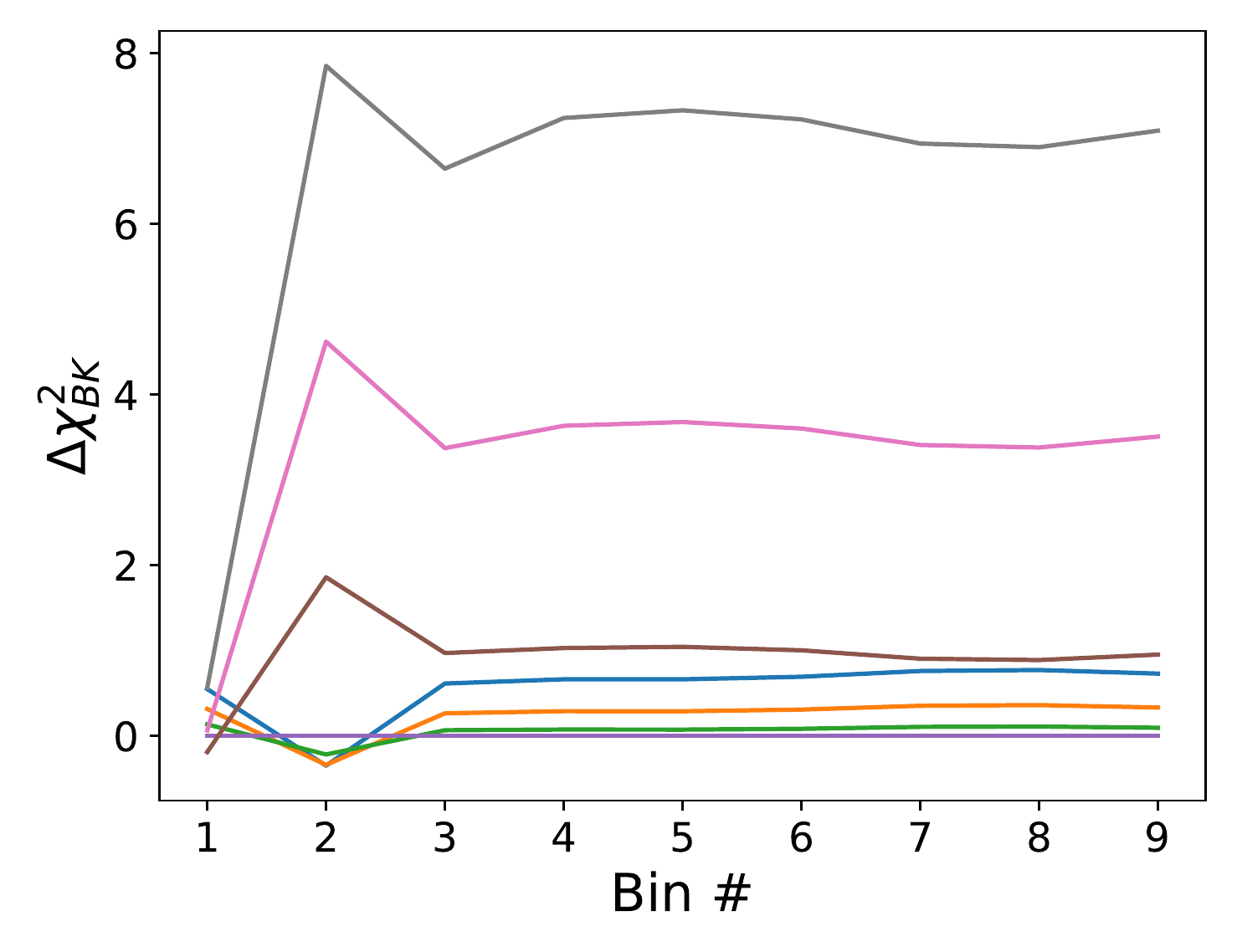}}
\caption{$\Delta\chi^2=\chi^2-\chi^2_{bestfit}$ for BK18 with data
cuts. The $x$-axis represents that only the first $x$ bins
(counting from low-$l$) out of the total nine $l$-bins of BK18 are
used in the calculation of $\chi^2$ and $y$-axis for the
difference in BK18 $\chi^2$ compared with the bestfit model with
the same data cuts. Complete BK18 dataset corresponds to bin \#=9
on the horizontal axis.}
    \label{chi2bin}
\end{figure}

\textbf{The difference in upper bound is related to the different
bestfit values of $r$ in $\Lambda$CDM and EDE.} In Fig.\ref{r-chi2}, we plot the response of total
BK18 $\chi^2$ to the variation in $r$ in $\Lambda$CDM and AdS-EDE, respectively. It is clear that the $\chi^2_{BK}$ responds
to $r$ above the bestfit point very similarly in both models, thus
the difference in their bestfit values are carried over to the
derived 95\% C.L. upper bounds, i.e. in Table.\ref{partab} the
difference in $r$ upper bounds between $\Lambda$CDM and EDE is
approximately equal to the difference in the corresponding bestfit
values for both EDE models.

\textbf{The increase of power in $C_{l,lensing}^{BB}$ (see the
dotted green and dashed black lines in Fig.\ref{clbb}) is a major
contributing factor to the difference in the bestfit values of $r$
between EDE and $\Lambda$CDM}. It remains to understand why such
distinct values of $r$, i.e. $r=0.015$ for $\Lambda$CDM and
$r=0.008$ for AdS-EDE, can fit the BK18 data equally well. To this
end, we perform a crude data cut to BK18 - using only part of
BK18's nine $l$-bins (corresponding to the nine data points in
Fig.\ref{clbb}) to calculate $\chi^2$. The results are plotted in
Fig.\ref{chi2bin}. Overall it is the first three bins ($l=37-120$)
that are most sensitive to the variation in $r$ since they
contribute most of the change in $\Delta\chi^2_{BK}$. The first
bin favors (i.e. smallest $\Delta\chi^2_{BK}$) non-zero $r$. On
the other hand, large $r$ lines peak at the second bin while the
line with the smallest $r=0.001$ (blue) displays the deepest dip
there, showing a preference for $r\lesssim0.001$. Bin \#3 on the other hand
prefers $r\gtrsim0.04$.

\begin{figure}
\subfigure[Different model]{\includegraphics[width=0.48\linewidth]{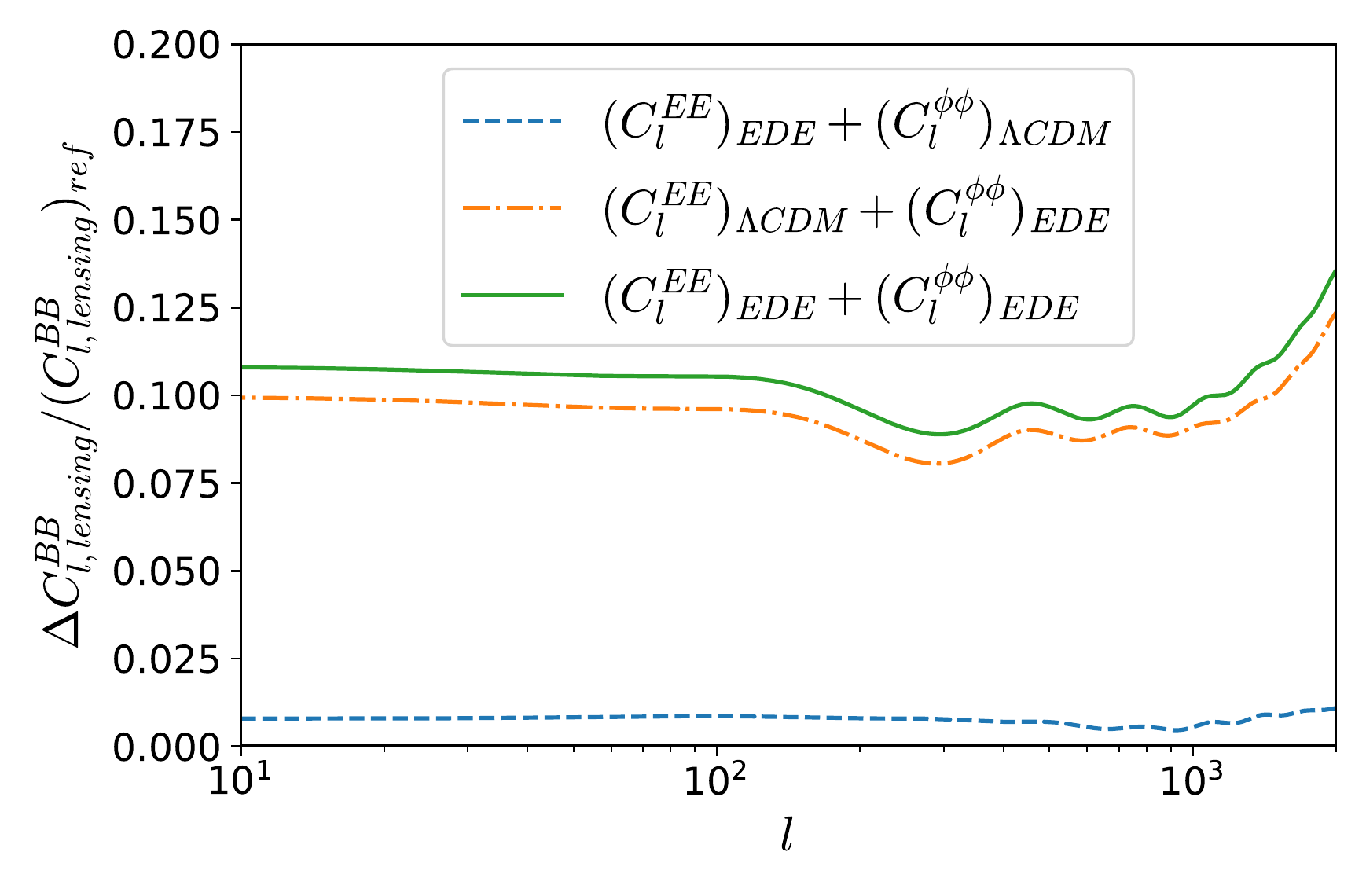}\label{lensBmod}}
\subfigure[Different $l$ range]{\includegraphics[width=0.48\linewidth]{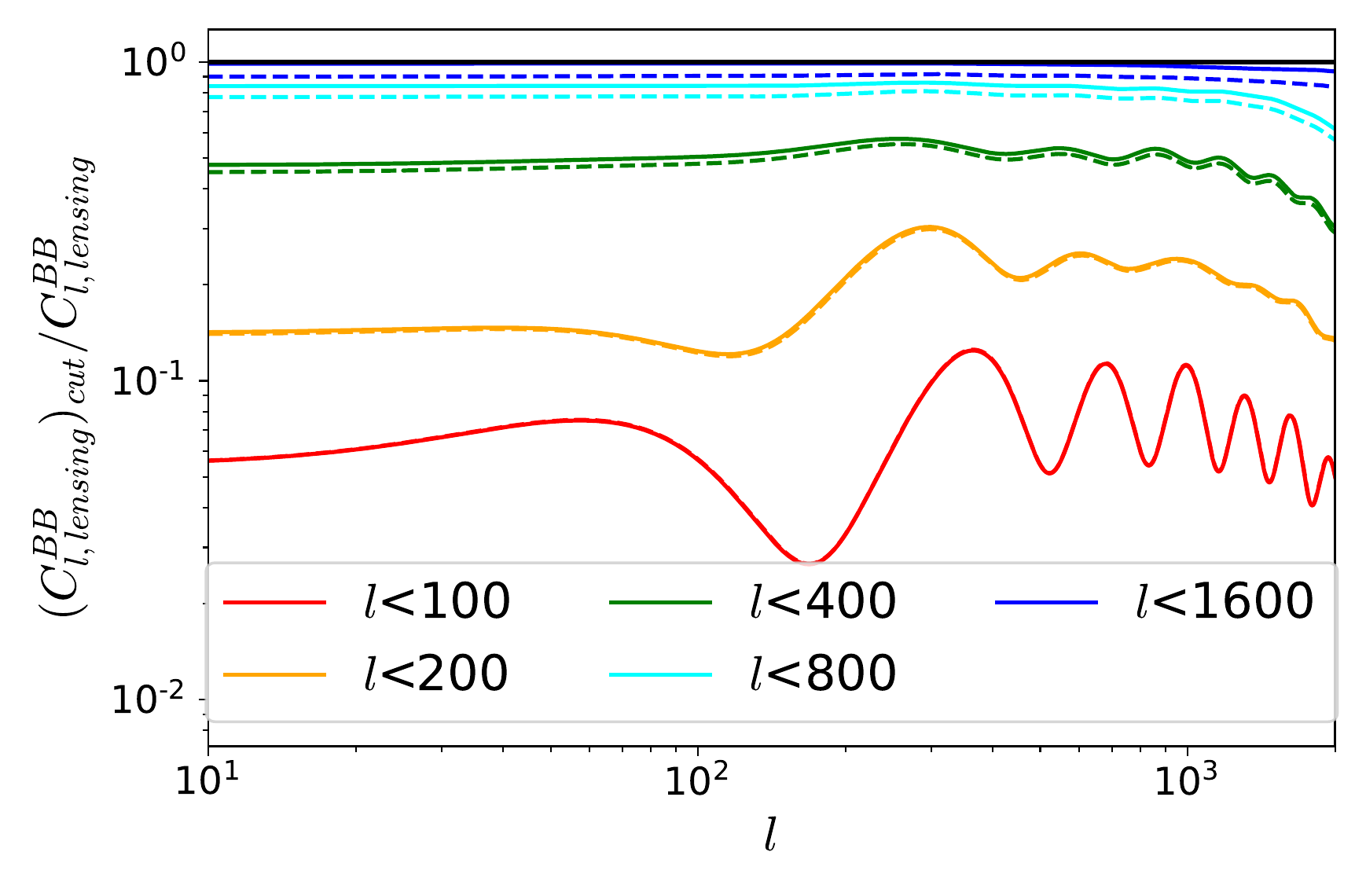}\label{lensBcut}}
\caption{The lensing B-mode calculated using different
$C_l^{\phi\phi}$ and $C_l^{EE}$. \textit{Left panel:} Relative
difference in $C_{l,lensing}^{BB}$ compared with $\Lambda$CDM
bestfit. Results are calculated using different combination of
unlensed $C_l^{EE}$ and $C_l^{\phi\phi}$ from the $\Lambda$CDM and
AdS-EDE bestfit model. \textit{Right panel:}
$C_{l,lensing}^{BB}$ calculated from bestfit AdS-EDE unlensed $C_l^{EE}$,
using different portion of the $C_l^{\phi\phi}$ from the same
model, compared with the full lensing $C_l^{BB}$.}
    \label{lensB}
\end{figure}

Bin \#3 ($l=91-120$) is much less sensitive (i.e. smaller increment
in $\Delta\chi^2_{BK}$ for $r<0.001$ in  Fig.\ref{chi2bin}) to
small $r$ in EDE\footnote{The same applies to bin \#1, but the
constraining power on $r$ comes mostly from bin \#3 due to its
higher data quality.}, which suggests that a much smaller $r$ is
allowed to fit bin \#2 better without significantly worsening the
fit at bin \#3. $C_{l,tot}^{BB}$ in EDE and $\Lambda$CDM are
actually well constrained by BK18 data near bin \#3, see the zoomed-in region
of Fig.\ref{clbb}, both bestfit lines very close to each other and
intersecting. \textbf{In the bestfit $\Lambda$CDM,
$C_{l,lensing}^{BB}$ contributes roughly 80\% power of the total
amplitude near $l=100$, thus the 10\% increase, see
Fig.\ref{lensB}, in $C_{l,lensing}^{BB}$ from $\Lambda$CDM to EDE
needs a 50\% reduction in $C_{l,tensor}^{BB}$ to keep
$C_{l,tot}^{BB}=C_{l,tensor}^{BB}+C_{l,lensing}^{BB}$
approximately unchanged, corresponding to a 50\% reduction in $r$
bestfit, i.e. $\delta r\sim 0.007$.} It is surprising that this
rather crude estimate explains the difference in bestfit values of
$r$ in Table.\ref{partab} pretty well.

\textbf{One of the sources of the enhanced $C_{l,lensing}^{BB}$ in
EDE is an excess of power in $C_l^{\phi\phi}$ at $l=200-800$, away
from its primary peak.} We compare the $C_{l,lensing}^{BB}$
brought by different
$C_l^{\phi\phi}$ from different unlensed $C_l^{EE}$ in Fig.\ref{lensB}. We set $A_{L}=1$ throughout this paper. It is clear from
Fig.\ref{lensBmod} that the enhancement in $C_{l,lensing}^{BB}$ nearly
entirely comes from the difference in $C_l^{\phi\phi}$. This is
understandable since $C_l^{EE}$ (lensed) in both bestfit models
are tightly constrained by the Planck data, and so nearly the
same. The major contribution to the difference in
$C_{l,lensing}^{BB}$ between $\Lambda$CDM and AdS-EDE comes from
$C_l^{\phi\phi}(200<l<800)$, see Fig.\ref{lensBcut}.
Interestingly, the peak of $C_l^{\phi\phi}$ (related to the peak
of the matter power spectrum $P_k$, corresponding to scales
entering horizon near matter-radiation equality) is excluded from
this multiple range. It also explains why in Fig.\ref{clbb}
changing $n_s$ has nearly no effect on the lensing B-mode, since
$200<l<800$ is near the pivot scale
$k_{pivot}=0.05\text{Mpc}^{-1}$ which is much more sensitive to
$A_s$ rather than $n_s$.

\begin{figure}
\includegraphics[width=0.8\linewidth]{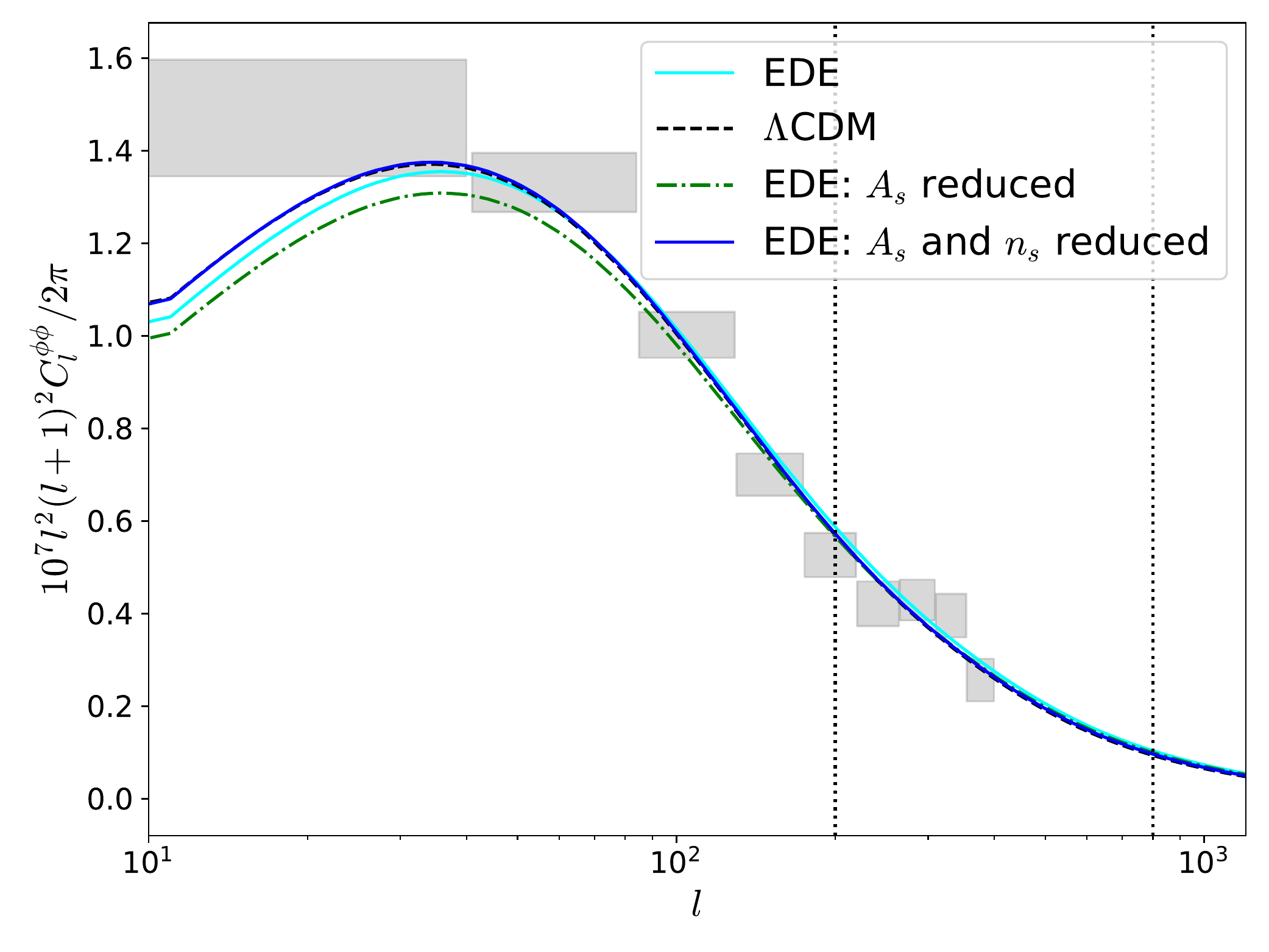}
\caption{The lensing convergence spectrum
$C_l^{\kappa\kappa}=l^2(l+1)^2C_l^{\phi\phi}/4$ in the bestfit
$\Lambda$CDM and AdS-EDE models. AdS-EDE models with $A_s$ and/or
$n_s$ set to their $\Lambda$CDM bestfit values are also plotted.
The vertical dotted lines mark the position of $l=200$ and
$l=800$. The gray shaded regions are the band power constraints
from Planck18 lensing reconstruction \cite{Planck:2018lbu}.
Comparing the solid blue and green dashed dotted lines, changing
$n_s$ affects the peak (away from pivot) but not the region
$200<l<800$ (near the pivot).}
    \label{clpp}
\end{figure}

\textbf{We attribute the additional power in
$C_l^{\phi\phi}(200<l<800)$ to the increment in $\omega_{cdm}$ and
$A_s$,} which enhances amplitude of the matter power spectrum
$P_k$ on all scales smaller than $k_{eq}$, the peak of $P_k$,
so $C_l^{\phi\phi}(200<l<800)$. On the other hand, for EDE, the increment in
$\omega_{cdm}$ in fact has minor effect on the peak (located near
$l\sim60$) height of $C_l^{\phi\phi}$, see appendix-\ref{apdx:omh}
for details. The shift (\ref{deltans}) of $n_s$
suppresses power at $l\sim60$ by $1-(l_{peak}/l_{pivot})^{\delta
n_s}\sim1-(60/500)^{0.03}\approx6\%$ but does not obviously affect
$200<l<800$ which is near $l_{pivot}\simeq500$. AdS-EDE brings nearly identical $C_l^{\phi\phi}$
near $l\lesssim100$ as $\Lambda$CDM does under the same initial
condition. To compensate for the power deficit at the lensing peak
$l\simeq60$ caused by increasing $n_s$, the amplitude $A_s$ has been slightly
increased in AdS-EDE, which further enhances $C_l^{\phi\phi}$ in
the relevant multiple range $200<l<800$. The above
observation about $n_s$ and $A_s$ has been confirmed in
Fig.\ref{clpp}.

\begin{figure}
\includegraphics[width=0.8\linewidth]{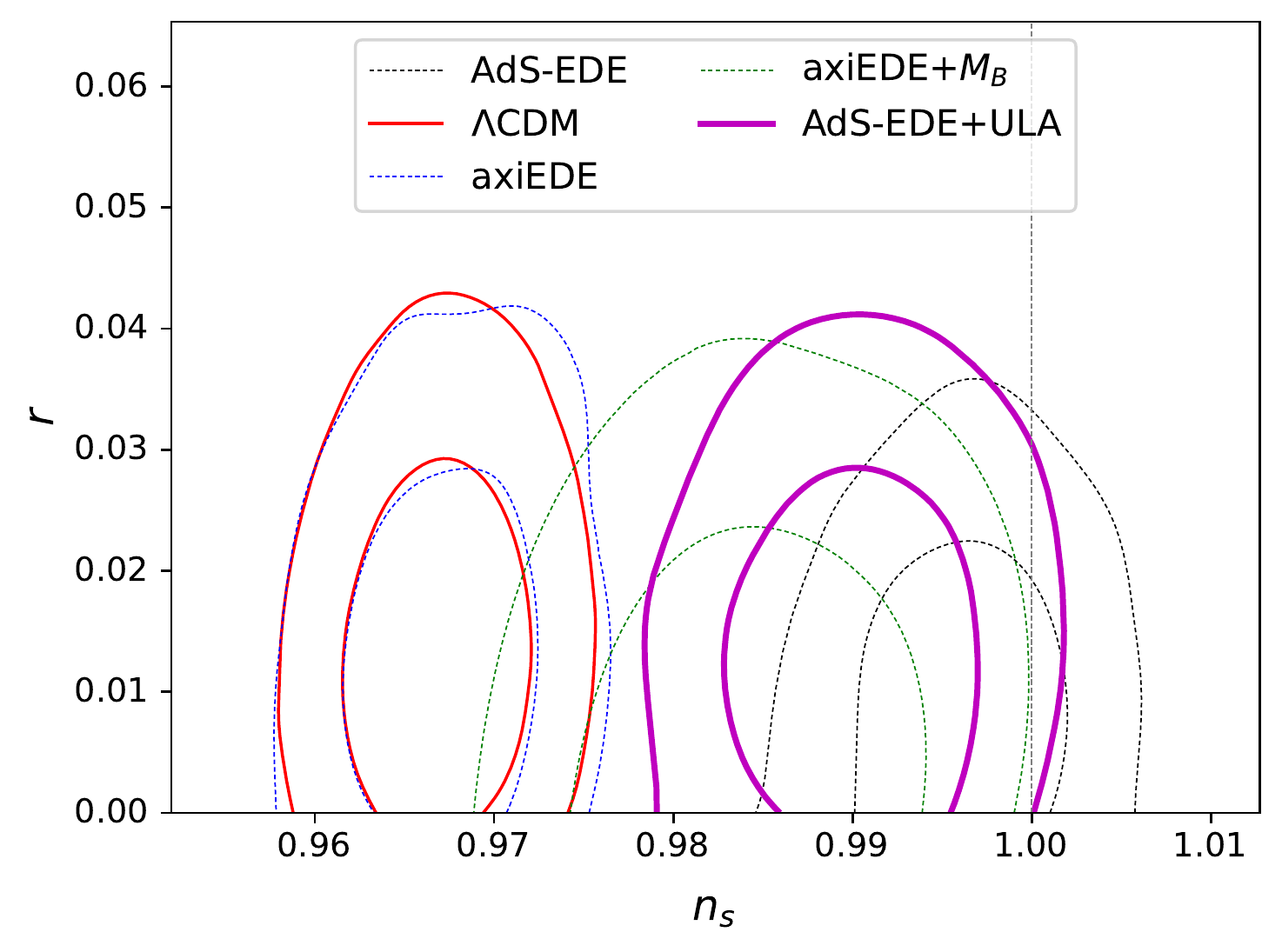}
\caption{$r-n_s$ plot highlighting the $\Lambda$CDM and AdS-EDE plus
ultra light axion (ULA) models. The upper bound on $r$ in AdS-EDE+ULA is
relaxed to be comparable with $\Lambda$CDM.}
    \label{r_ns_axi}
\end{figure}

As clarified, it is the enhanced matter power spectrum, so lensing
potential $C_{l,lensing}^{BB}$, at $l>200$ that results in the
tightened bound on $r$. Thus suppressing matter power spectrum on
small scales will possibly relax the bound, which might be
relevant with resolving $S_8$ tension. As a cross check, we
confront the AdS-EDE plus ultra light axion model
\cite{Ye:2021iwa} (able to restore cosmological concordance with
both $S_8$ and $H_0$) with P18+BK18+BAO+SN dataset as well as the
Gaussian prior $S_8=0.755^{+0.019}_{-0.021}$
\cite{Asgari:2019fkq}, and plot the MCMC results in
Fig.\ref{r_ns_axi}. As expected, the constraints on $r$ is relaxed
to $r<0.034 \ (95\%\text{C.L.})$ (but still $n_s\approx 1$),
comparable to the $\Lambda$CDM result $r<0.035 \
(95\%\text{C.L.})$.

\section{Conclusion}\label{sec:conclusion}

We present the first constraint on primordial GWs, quantified as
the tensor-to-scalar ratio $r$, in Hubble-tension-free EDE
cosmologies using the most recent BK18 data. It is found that the
upper bound on $r$ gets tightened in correlation with the
increment in $H_0$, and the most stringent bound $r<0.028$, as
opposed to $r<0.036$ reported by the BICEP/Keck collaboration for
$\Lambda$CDM \cite{BICEP:2021xfz}, is obtained using the EDE model
with the largest Hubble constant, i.e. $H_0=72.36^{+0.49}_{-0.56}$
km/s/Mpc in AdS-EDE. We argued that this tightening of bound is a
manifestation of the competition between $C^{BB}_{l,tensor}$ and
$C^{BB}_{l,lensing}$ with
$C^{BB}_{l,tot}=C^{BB}_{l,tensor}+C^{BB}_{l,lensing}$ constrained
by the BK18 data. In the EDE models, the increment in
$\omega_{cdm}$ and $A_s$ brings more power at intermediate and
small scales ($l>200$) in the matter power spectrum, enhancing
$C_l^{\phi\phi}(200<l<800)$ and thus $C^{BB}_{l,lensing}$. As a
consequence, the $C^{BB}_{l,tensor}$ allowed by BK18 must be
lowered, so the smaller $r$.

Our results underline that the exact upper bound on $r$ is
model-dependent even with the most recent BK18 data. The take-away
message for primordial Universe model building is that though the
order of magnitude constraint from observation
$r\lesssim\mathcal{O}(10^{-2})$ is to be respected, detailed value
of the bound might be different in different cosmological models,
which will have profound implication to inflation and early
Universe physics. Our results also highlight the crucial roles of
weak lensing from scales
$k=\mathcal{O}(0.01\sim0.1\text{Mpc}^{-1})$, probed by galaxy
surveys such as DES \cite{DES:2021wwk} and Euclid
\cite{Euclid:2021qvm}, in constraining $r$ using BICEP/Keck data, and of accurate measurements of CMB on small scales, with surveys
such as Simons Observatory \cite{SimonsObservatory:2018koc} and
CMB-S4 \cite{CMB-S4:2016ple}, in differentiating the
Hubble-tension-free cosmologies. It might also be interesting to restudy $A_L$ with the BK18 data \cite{BICEP:2021xfz} in beyond $\Lambda$CDM cosmologies which modify $C_l^{\phi\phi}$.

\paragraph*{Acknowledgments}
This work is supported by the NSFC, No.12075246, the KRPCAS,
No.XDPB15. We acknowledge the Tianhe-2 supercomputer for providing
computing resources. Some figures are generated using GetDist
\cite{Lewis:2019xzd}. We thank Alessandra Silvestri for useful comments and discussions.

\appendix
\section{More MCMC results}\label{apdx:mcmc}

\begin{figure}[h]
	\includegraphics[width=\linewidth]{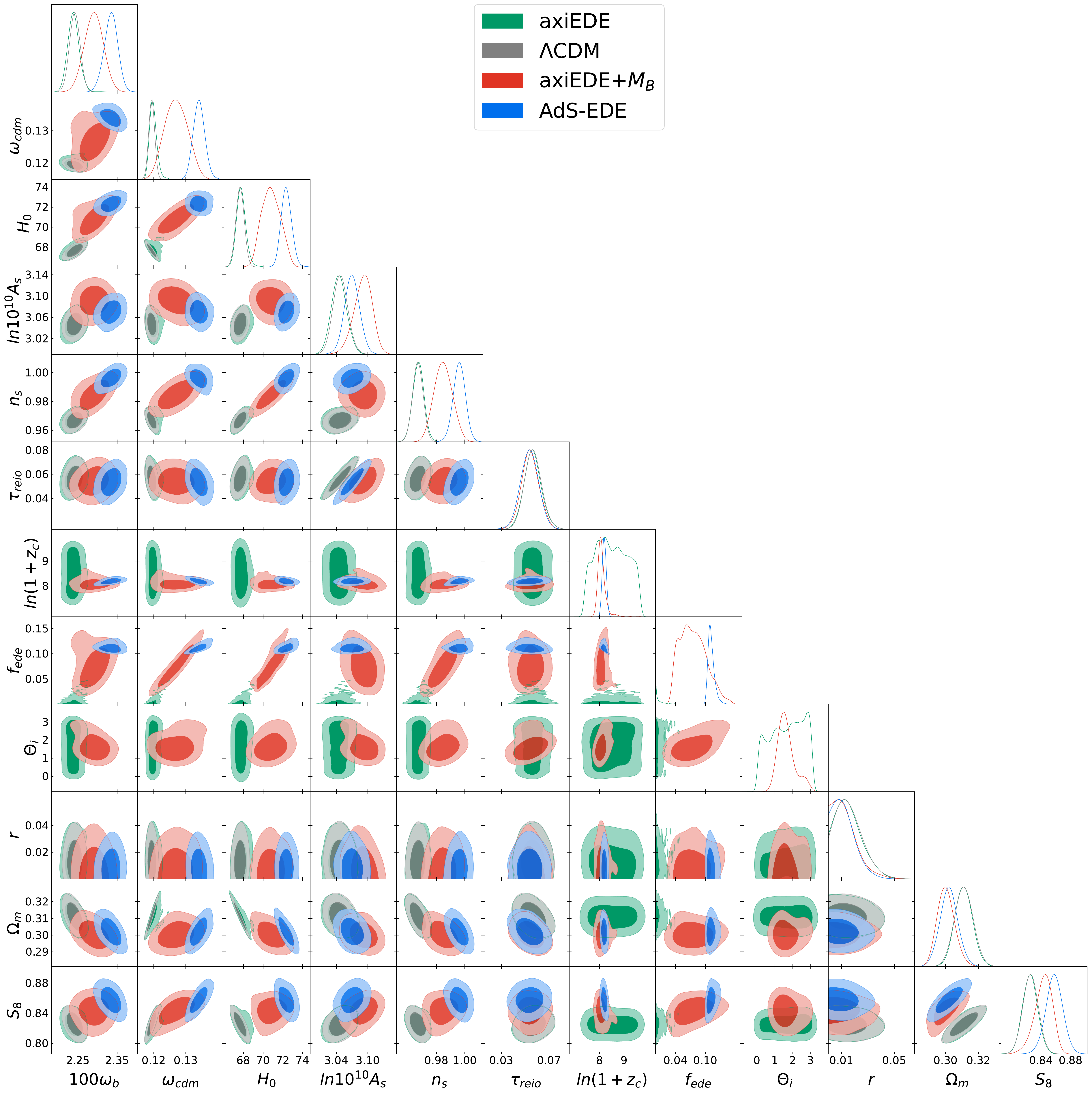}
	\caption{68\% and 95\% posterior distributions of all cosmological
		parameters in the $\Lambda$CDM and EDE models.}
	\label{triangle-full}
\end{figure}
\begin{table}
	\begin{tabular}{|c|c||c|c|c|}
		\hline
		Dataset&$\Lambda$CDM&axiEDE (w/o $H_0$ prior)&axiEDE (w/ $H_0$ prior)&AdS-EDE\\
		\hline
		Planck high-$l$ TTTEEE&2349.64 &2348.61 &2347.00 &2348.16 \\
		Planck low-$l$ TT&23.61 &21.09 &21.06 &20.54 \\
		Planck low-$l$ EEBB&395.90 &394.09 &392.77 &392.74 \\
		Planck lensing&8.95&9.60&9.91&11.17\\
		BK18&537.25 &536.70 &536.01 &535.93 \\
		BAO&5.78 &5.24 &5.62 &5.46 \\
		SN&-- &1027.04 &1026.86 &1026.86 \\
		$M_B$ prior&-- &-- &4.65 &-- \\
		\hline
	\end{tabular}
	\caption{Bestfit $\chi^2$ for each likelihood. The $\Lambda$CDM
		bestfit is taken to be the point with the lowest total $\chi^2$
		value in the publicly available chains by BK18.}
	\label{chi2}
\end{table}

Table.\ref{partab} and Fig.\ref{triangle-full} show the posterior
results for cosmological parameters. Despite using exactly the
same MCMC chains as BK18, we obtained a slightly smaller upper
bound $r<0.035$, which we attribute to different analysis
configuration in the GetDist package. Table.\ref{chi2} presents
the per experiment bestfit $\chi^2$ for each model.

As noted in earlier works, without a $H_0$($M_B$) prior, axiEDE
with Planck CMB data on its own yields nearly identical results to
$\Lambda$CDM and $f_{ede}$ is compatible with zero, despite a
bestfit point outside the 1$\sigma$ contours where EDE is
non-negligible and $H_0$ is much larger. This is because both the
$\Lambda$CDM bestfit ($f_{ede}=0$) and the EDE bestfit
($f_{ede}\ne0$) are local minima in the full phase space, but
other EDE parameters (such as $\Theta_i$ or $z_c$) are essentially
free in the $f_{ede}=0$ case, thus the phase space around the
$\Lambda$CDM bestfit viable to the MCMC chain is dimensionally
larger than that around the EDE bestfit point. And without any
$H_0$($M_B$)-ralated prior, the EDE bestfit is only marginally
better than the $\Lambda$CDM one, see $\chi^2$ in axiEDE in
Table.\ref{chi2}, thus the MCMC chain will inevitably center
around the $\Lambda$CDM local minima, see also
\cite{Herold:2021ksg} for recent discussion.

The AdS-EDE model resolves the Hubble tension even without a
$H_0$($M_B$)-related prior partially because the AdS phase masks
out the $\Lambda$CDM bestfit point since too small $f_{ede}$ will
result in the field confined in the disastrous AdS region and is
not favored.

\section{CMB lensing constraints in AdS-EDE}\label{apdx:omh}
\begin{figure}
\includegraphics[width=0.8\linewidth]{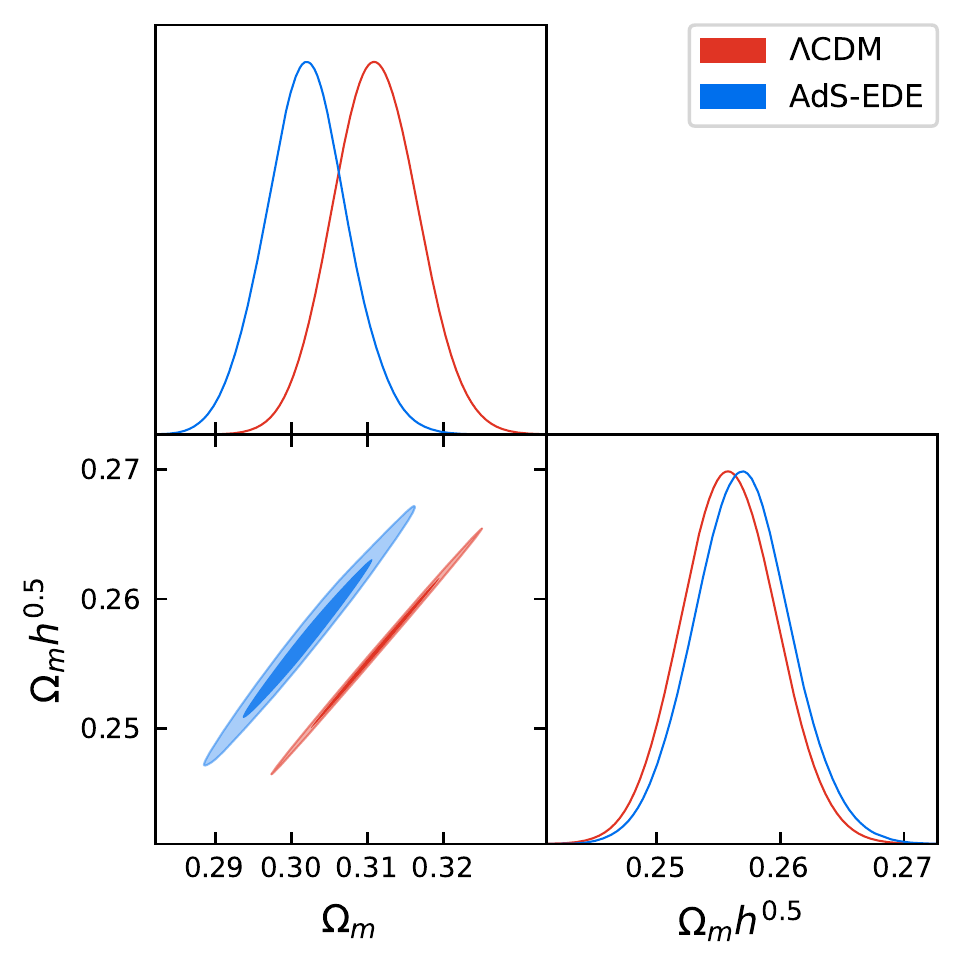}
\caption{Posterior distribution of $\Omega_m$ versus
$\Omega_mh^{0.5}$.}
    \label{omh}
\end{figure}
As can be seen in Fig.\ref{clpp}, $C_l^{\phi\phi}$ is nearly
identical in both AdS-EDE and $\Lambda$CDM around its peak under
the same initial conditions (i.e. $A_s$ and $n_s$). This is
because in EDE the parameter most relevant to the peak height of
$C_l^{\phi\phi}$ is $\Omega_mh^{0.5}$ rather than
$\omega_m=\Omega_mh^2$. Using the variable $y=h\chi$, $\chi$ being
the angular diameter distance, and the Limber approximation, the
lensing convergence spectrum
$C_l^{\kappa\kappa}=l^2(l+1)^2C_l^{\phi\phi}/4$ writes
\begin{equation}\label{clkk}
    C_l^{\kappa\kappa}=\frac{9}{4}\Omega_m^2h\int_{0}^{\infty}dy a^{-2}(y)\hat{g}_L^2(y)P_m\left(k=\frac{l}{\chi}h,\eta(y)\right)
\end{equation}
with $\eta$ the conformal time. The lensing kernel $\hat{g}_L$
with the window function $W(y)$ is
\begin{equation}
    \hat{g}_L(y)=\int_y^\infty
    dy'\left(1-\frac{y}{y'}\right)W(y').
\end{equation}
The angular location of the matter power
spectrum peak ($l_{eq}$) is constrained by CMB data in EDE
\cite{Poulin:2018cxd} as well as its peak height. Thus the peak
height of $C_l^{\kappa\kappa}$ is relevant to the prefactor $\Omega_mh^{0.5}$ in \eqref{clkk}.
The Planck lensing reconstruction therefore constrains $\Omega_m h^{0.5}\sim const.$, which is nearly invariant across
both models in Fig.\ref{omh}, while $\Omega_m$ show some
difference. Note this observation does not contradict the
result of Ref.\cite{Ye:2020oix}, in which $\Omega_m\sim const.$ is
the background level CMB+BAO constraint. In the actual analysis,
since there is residual freedom in both data constraints, the MCMC results
make a compromise between different datasets. Thus the actual
degeneracy direction is
\begin{equation}
    \Omega_m\sim h^{-\alpha},\quad  0<\alpha<0.5,
\end{equation}
or equivalently $\omega_m\sim h^\beta$ with $1.5<\beta<2$. As an
example, numerical principle component analysis of the AdS-EDE
chain yields $\Omega_mh^{0.4}\sim const.$, so $\alpha=0.4$ and
$\beta=1.6$.

\bibliography{ref}
\end{document}